\documentclass[%
 reprint, superscriptaddress, amsmath,amssymb, aps,  prl]{revtex4-2}

 \pdfoutput=1

\usepackage{graphicx}%
\usepackage{dcolumn}%
\usepackage{bm}
\usepackage{color} 

\begin{document}


\title{Low-energy collective oscillations and Bogoliubov sound in an exciton-polariton condensate}

\author{E.~\surname{Estrecho}}
\email{eliezer.estrecho@anu.edu.au}
\author{M.~\surname{Pieczarka}}
 \email{Currently at Department of Experimental Physics, Wroc\l aw University of Science and Technology, Wyb. Wyspia\'{n}skiego 27, 50-370 Wroc\l aw, Poland}
\author{M.~\surname{Wurdack}}
 \affiliation{ARC Centre of Excellence in Future Low-Energy Electronics Technologies \& Nonlinear Physics Centre, Research School of Physics,
The Australian National University, Canberra ACT 2601, Australia}
\author{M.~\surname{Steger}}
 \email{Currently at National Renewable Energy Lab, Golden, CO 80401, USA}
 \affiliation{Department of Physics and Astronomy, University of Pittsburgh, Pittsburgh, PA 15260, USA}
\author{K.~\surname{West}}
\author{L.~N.~\surname{Pfeiffer}}
 \affiliation{Department of Electrical Engineering, Princeton University, Princeton, NJ 08544, USA}
\author{D.~W.~\surname{Snoke}}   
 \affiliation{Department of Physics and Astronomy, University of Pittsburgh, Pittsburgh, PA 15260, USA}
\author{A.~G.~\surname{Truscott}}
 \affiliation{Laser Physics Centre, Research School of Physics,
The Australian National University, Canberra ACT 2601, Australia}
\author{E.~A.~\surname{Ostrovskaya}}
 \email{elena.ostrovskaya@anu.edu.au}
 \affiliation{ARC Centre of Excellence in Future Low-Energy Electronics Technologies \& Nonlinear Physics Centre, Research School of Physics,
The Australian National University, Canberra ACT 2601, Australia}
\date{\today}

\begin{abstract}

We report the observation of low-energy, low-momenta collective oscillations of an exciton-polariton condensate in a round ``box'' trap. The oscillations are dominated by the dipole and breathing modes, and the ratio of the frequencies of the two modes is consistent with that of a weakly interacting two-dimensional trapped Bose gas. The speed of sound extracted from the dipole oscillation frequency is smaller than the Bogoliubov sound, which can be partly explained by the influence of the incoherent reservoir. These results pave the way for understanding the effects of reservoir, dissipation, energy relaxation, and finite temperature on the superfluid properties of exciton-polariton condensates and other two-dimensional open-dissipative quantum fluids.

\end{abstract}

\maketitle
{\em Introduction.---} Low-energy collective excitations directly probe Bogoliubov sound in quantum fluids and gases, which in turn provides critical information on the thermodynamic and superfluid properties of these systems. Propagation of sound in two-dimensional (2D) quantum fluids~\cite{Ville2018,Ota2018,Ota2018b,Cappellaro2018,Bohlen2020} is particularly interesting, since it is governed by Berezinskii--Kosterlitz--Thouless (BKT) rather than Bose-Einstein condensation (BEC) physics. Furthermore, collective excitations in 2D quantum gases can reveal quantum corrections to classical symmetries~\cite{Peppler2018,Holten2018}, and quantum phase transitions ~\cite{Tanzi2019,Natale2019,Guo2019}.

2D non-equilibrium condensates of exciton polaritons (polaritons hereafter)~\cite{Deng2002,Kasprzak2006,Balili2007,Deng2010,Carusotto2013,Byrnes2014} exhibit a wide range of phenomena including BKT~\cite{Roumpos2012,Nitsche2014,Caputo2018} and Bardeen--Cooper--Schrieffer (BCS)~\cite{Keeling2005, Byrnes2010,Hu2019} physics, as well as superfluid-like behavior~\cite{Juggins2018,Nardin2011,Amo2009,Lerario2017}. Their collective excitation spectrum is complex, and is expected to differ from the Bogoliubov prediction for an equilibrium BEC, especially at small momenta, due to dissipation~\cite{Keeling2004,Wouters2007}, coupling to an incoherent excitonic reservoir~\cite{Byrnes2012,Smirnov2014} and finite-temperature effects. Contrary to expectations, recent experiments~\cite{Utsunomiya2008,Stepanov2019,Pieczarka2020,Ballarini2020} revealed an excitation spectrum consistent with Bogoliubov theory. However, these experiments either could not probe the low-momenta region of the excitation spectrum \cite{Stepanov2019,Pieczarka2020}, or were performed in a near-equilibrium regime \cite{Ballarini2020} without the influence of the reservoir. Hence, direct access to the low-energy excitations and their damping rates is needed to understand the influence of nonequilibrium and finite temperature effects on polariton superfluid dynamics.

\begin{figure}
\includegraphics[width=\columnwidth]{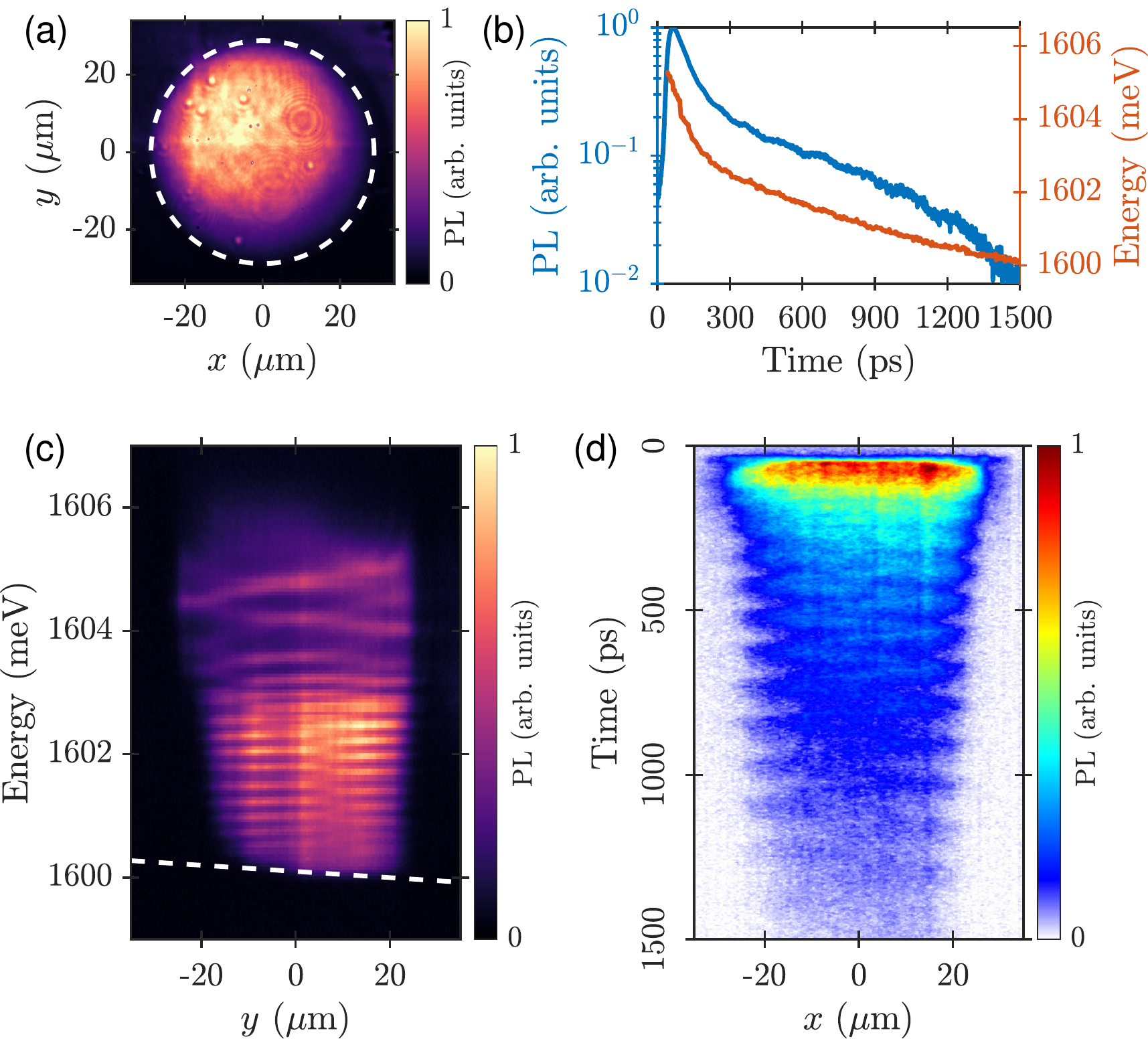}
\caption{\label{fig1} Sloshing interaction-dominated polariton condensate. (a)~Time-integrated real space image of the condensate trapped in a round box potential (dashed line) of the diameter $D=56~\mu$m.
	(b) Time-resolved, spatially integrated, normalized PL intensity (blue) and the condensate energy (orange) measured at the center of the trap.
	(c) Time-integrated real space spectrum of the condensate along the $y$-axis at $x\approx 0$. Dashed line is a guide to the eye indicating the tilted bottom of the optically-induced potential.
	(d) Time-resolved real space distribution of the condensate along the $x$-axis at $y\approx 0$.}
\end{figure}

In this Letter, we report the observation of low-energy collective oscillations of a trapped 2D polariton condensate. Using an optically-induced round box trap in the pulsed excitation regime~\cite{Estrecho2019}, we create an interaction-dominated condensate undergoing long-lived ``sloshing'' (Fig.~\ref{fig1}).
The frequency ratio of the two normal modes of the dynamics, i.e. center-of-mass (dipole) and breathing (monopole) oscillations, is in remarkable agreement with the theory for a weakly interacting 2D Bose gas in thermal equilibrium.
However, the speed of sound extracted from the dipole mode frequency is lower than expected and has a non-trivial dependence on the interaction energy, pointing to a strong influence of the incoherent reservoir on the low-energy collective excitations. 

{\em Experiment.---}The polaritons are created in a high-quality GaAs/AlAs $3\lambda/2$ microcavity sandwiched between 32(top) and 40(bottom) pairs of distributed Bragg reflectors with $12$ embedded GaAs quantum wells of nominal thickness $7$~nm~\cite{Nelsen2013}.
A high-density condensate is formed in an optically induced round trap [see Fig.~\ref{fig1}(a)] using an off-resonant pulsed excitation in a geometry similar to Ref.~\cite{Estrecho2019}. A spatial hole-burning effect~\cite{Estrecho2018, Estrecho2019} ensures that the pump-induced trapping potential is box-like, which is reflected by the sharp edges of the condensate density distribution in the Thomas-Fermi regime, see Supplemental MateriaI (SM). Time-resolved spectral imaging of real space (RS) and $k$-space (KS) dynamics is enabled by a streak camera. The condensate forms in ${\sim}50$~ps and then slowly decays, resulting in a fast rise and slow fall of the time-resolved photoluminescence (PL) signal, as shown in Fig.~\ref{fig1}(b). Decay of the condensate density leads to decreasing energy blueshift $\Delta E$ associated with the condensate, which results in one-to-one correspondence between time and the condensate energy. 

The condensate usually decays on a timescale of the polariton lifetime (${\sim}200$ ps), which makes it {\em impossible} to track the slow dynamics of collective oscillations. To overcome this limitation, we use a high-energy photo-excitation ${\sim}150$~meV above the lower polariton resonance, which produces a large reservoir with a very low effective decay rate~\cite{Anton2013} (see SM). The lifetime of the condensate replenished by this reservoir is much greater than the polariton lifetime, as evidenced by the long PL decay time (up to $1.5$~ns) [Fig.~\ref{fig1}(b)]. This also results in a bright low-energy tail in the RS spectrum [see Fig.~\ref{fig1}(c)]. Slow decay of the condensate is key to our measurement.

{\em Results.---} Time-integrated imaging of the condensate [e.g., Fig.~\ref{fig1}(a)] typically washes out all dynamics. In this work, due to the well-defined time-resolved energy of the condensate [see Fig.~\ref{fig1}(b)], the energy-resolved RS distribution shown in Fig.~\ref{fig1}(c) displays modulations of the PL intensity, indicating underlying density oscillations. Indeed, the time-resolved RS distribution, shown in Fig.~\ref{fig1}(d), reveals the spatial density oscillations, while the KS distribution shows that the majority of the polaritons occupy the $k\sim 0$ state (see SM).

\begin{figure}
	\includegraphics[width=\columnwidth]{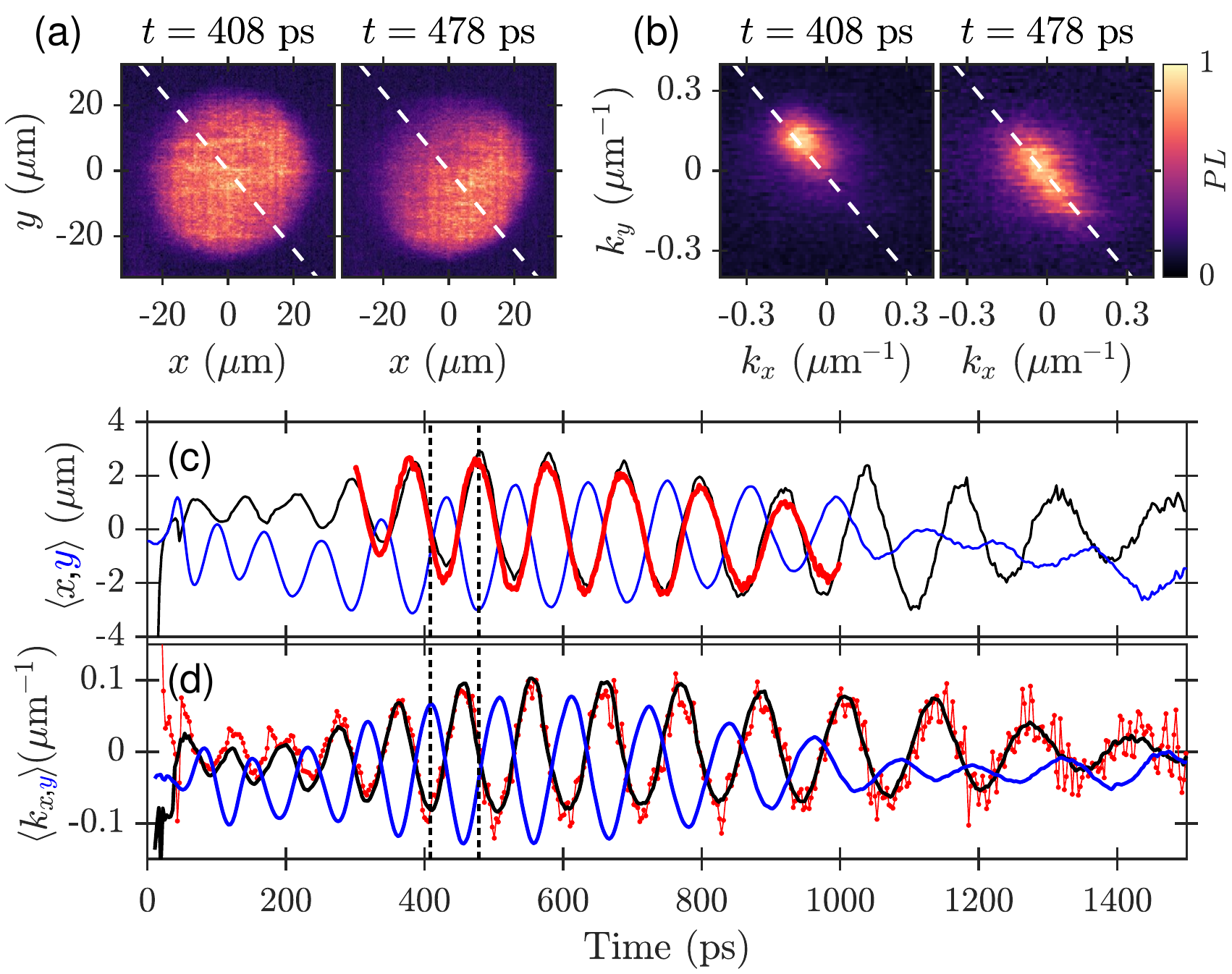}
	\caption{\label{fig2} 
	Snapshots of (a) RS and (b) KS condensate density distributions at two different times.
	Dashed lines show the direction of the microcavity wedge.
	Time-evolution of the average (c) position $\langle x \rangle$ (black), $\langle y \rangle$ (blue) and (d) momentum $\langle \hbar k_x \rangle$ (black), $\langle \hbar k_y \rangle$ (blue), extracted from the condensate density distributions.
	Red line in (c) corresponds to the time-dependent amplitude $w(t)$ of the dipole mode in Fig.~\ref{fig3}(b).
	Red dots in (d) correspond to $(m/\hbar) d\langle x \rangle/dt$ calculated from (c).
	Vertical dashed lines correspond to the snapshots in (a,b).
	}
\end{figure}

It is important to stress that the images in Fig.~\ref{fig1}(c,d) are accumulated over millions of condensation realisations. The persistent density modulations mean that the dynamics recurs despite the stochastic nature of the condensate formation in every experimental realisation~\cite{Estrecho2018}. This recurrence is due to a significant wedge of our microcavity (effective linear potential) \cite{Nelsen2013} oriented along the diagonal $x$-$y$ direction, as evidenced by the off-centred RS image in Fig.~\ref{fig1}(a) and the tilted low-energy tail of the energy-resolved RS distribution in Fig.~\ref{fig1}(c).

To analyze the observed oscillations, we perform time-resolved tomography on the RS $n_{r}(x,y,t)$ and KS $n_{k}(k_x,k_y,t)$ density distributions (see SM). The snapshots at different times [Fig.~\ref{fig2}(a,b)] reveal ``sloshing'' of the condensate along the cavity wedge (see SM for movies). When most of the polaritons are on one side of the trap, $n_k$ is centered at $k\approx0$ [right panels in Fig.~\ref{fig2}(a,b)], corresponding to zero average momentum at the classical turning point of the confining potential. When the $n_r$ is symmetric, $n_k$ is peaked [left panels in Fig.~\ref{fig2}(a,b)], corresponding to a large average momentum. This harmonic motion is summarized in Fig.~\ref{fig2}(c,d), where the expectation values of position $\langle x \rangle$ and momentum $\langle \hbar k \rangle$, are calculated as: $\langle x \rangle = \int x n_{r}dxdy/ \int n_{r}dxdy$ and $\langle k \rangle = \int k n_{k}dk_x dk_y/ \int n_{k}dk_x dk_y$.
The behavior of the expectation values is in excellent agreement with the Ehrenfest's theorem of quantum mechanics:
$
m \frac{d \langle x \rangle}{dt} = \langle \hbar k_x \rangle,
$
as demonstrated by the overlap of $\langle  k_x \rangle$ (solid black line) and $(m/\hbar) d\langle  x \rangle/dt$ (red dots) in Fig.~\ref{fig2}(d). Here, $m$ is the polariton effective mass at a near-zero energy detuning between the exciton and the cavity photon (see SM).

\begin{figure}
	\includegraphics[width=\columnwidth]{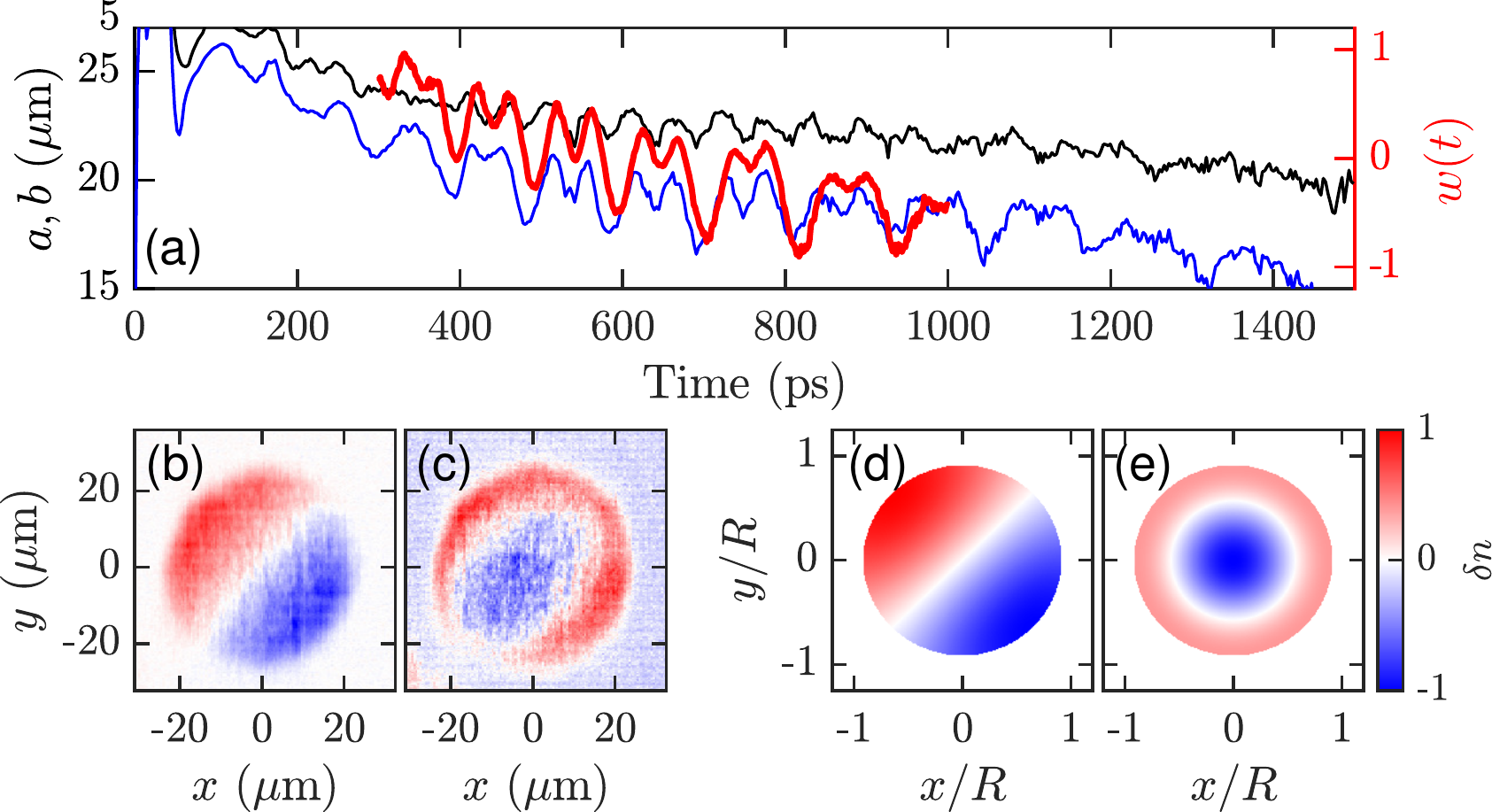}
	\caption{\label{fig3} 
	Time-evolution of the semi-axes of the ellipse $a$ (black) and $b$ (blue) fitted to the condensate density distribution $n_r(x,y,t)$. (b,c) Normal modes extracted by PCA. Red line in (a) corresponds to the amplitude $w(t)$ (in arbitrary units) of the PCA mode shown in (c). The amplitude $w(t)$ for the mode in (b) is shown in Fig.~\ref{fig2}(c). (d) Dipole and (e) breathing modes calculated from theory.}
\end{figure}

The condensate density distribution obtained from the experimental RS images indicates that the box trap has an elliptical rather than circular cross-section, with an aspect ratio of $1.2$. By fitting the edges of the density distribution with an ellipse with
semi-axes $a,b$ (see SM), we extract the time evolution of the condensate width shown in Fig.~\ref{fig3}(a). The oscillation frequency is clearly distinct from the average position or momentum in Fig.~\ref{fig2}(c,d).
To identify the normal modes of this oscillation, we employ a principal component analysis (PCA). This model-free statistical analysis tool~\cite{Segal2010, Dubessy2014,Natale2019} allows us to extract the spatial profile of the normal modes (principal components) $\mathcal{P}_c (x,y)$ and their time-dependent amplitudes $w_c(t)$ from the density $n_r(x,y,t) = \overline{n_r(x,y)} + \sum_{c=1}^{N_t} w_{c}(t) \mathcal{P}_c (x,y)$, where $\overline{n_r(x,y)}$ is the time average (see SM). The PCA reveals two dominant modes shown in Fig.~\ref{fig3}(b,c). The respective amplitudes $w(t)$, marked by the red line in Fig.~\ref{fig2}(c) and Fig.~\ref{fig3}(a), match the oscillations of the average position and width of the condensate.

The observed dynamics can be understood as a result of collective excitations of the polariton condensate. In the homogeneous, steady-state regime, polariton decay and interaction with the reservoir strongly modifies the dispersion of excitations at low momenta $k<k_s$~\cite{Keeling2004, Wouters2007}, where $k_s$ is determined by the polariton lifetime (see SM). For $k>k_s$, the dispersion recovers its sonic character, i.e. $\omega \approx ck$, where $c$ is the speed of sound. The long polariton lifetime ($\sim200$~ps) in our experiment results in a very small $k_s\approx0.01~\mu\mathrm{m}^{-1}$, an order of magnitude smaller than the momenta corresponding to the observed oscillations. For the sake of simplicity and in the absence of a better suited theory, we analyze the oscillations in the framework of equilibrium quantum hydrodynamics.

The total condensate density can be approximated by $n(x,y,t)=n_0 (x,y)+\delta n(x,y,t)$ where $n_0$ is the ground state density and $\delta n\ll n$ is the density of excitations. This approximation is validated by the low occupation of the normal modes relative to the condensate, $\delta N/N \sim 0.1$, extracted from PCA. In a round box trap of radius $R$, the density $n_0$ has a flat-top profile in the Thomas-Fermi limit~\cite{Estrecho2019}. Neglecting the sharp edges of $n_0$, the hydrodynamic equation for the collective excitations can be written as~\cite{Stringari1996} 
$\omega^2 \delta n=-c^2 \nabla^2 \delta n$. The wave equation is subject to the boundary conditions $\nabla \delta n=0$ at the edge ($r=R$) and the continuity condition in the azimuthal direction.
The normal modes have the form:
\begin{equation}
 \delta n_{l,m} \propto J_m \left(\frac{q_{l,m} r}{R} \right)e^{im\phi},
\end{equation}
where $J_m$ is the Bessel function of the 1st kind and $q_{l,m}$ is the $l$-th root of its derivative $J'_m$.
The indices $l,m$ denote the radial nodes and the orbital angular momentum of the mode, respectively, resulting in the dispersion:
\begin{equation}
\label{eq:dispersion_law}
\omega_{l,m} = c q_{l,m}/{R}.
\end{equation}

Of particular interest are the dipole ($l{=}1,m{=}\pm 1$) and the breathing ($l{=}1,m{=}0$) modes with the spatial profiles shown in Fig.~\ref{fig3}(d,e) (see SM for other modes) and frequencies $\omega_D\approx 1.84 c/R$  and $\omega_B\approx 3.83 c/R$, respectively. The dipole mode is a ``vortex''-like center-of-mass motion around the trap center but the reduced symmetry of the elliptical trap results in oscillations along the short (long) axis of the trap at a slightly higher (lower) frequency. The breathing mode shown in Fig.~\ref{fig3}(e) becomes a mixture of the monopole and quadrupole modes, with a pronounced oscillation of the condensate width that does not affect its center of mass. The dominant modes extracted by PCA [Fig.~\ref{fig3}(b,c)] are now readily identified as the dipole and breathing modes by comparison with Fig.~\ref{fig3}(d,e), and by matching the respective  amplitudes with the observed oscillations in Fig.~\ref{fig2}(c) and Fig.~\ref{fig3}(a). 

\begin{figure}
	\includegraphics[width=\columnwidth]{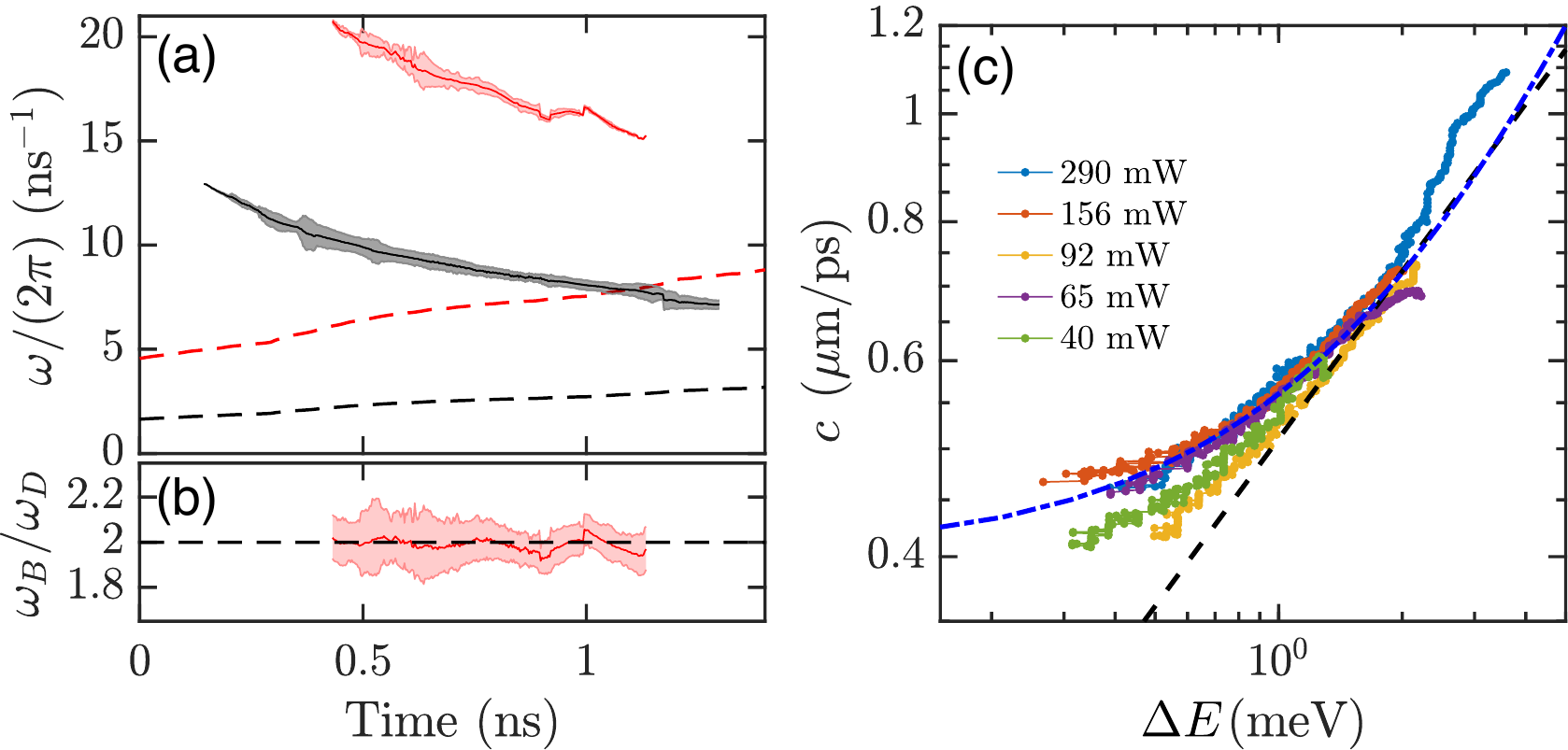}
	\caption{\label{fig4} Measurement of the speed of sound.
	(a) Frequency of the dipole (black) and breathing (red) modes. Black (red) dashed line corresponds to the non-interacting dipole (breathing) frequency $\omega^{\rm{sp}}_D$ $\left(\omega^{\rm{sp}}_B\right)$.
	(b) Ratio of the breathing and dipole frequencies $\omega_B/\omega_D$. Shaded areas in (a,b) correspond to the width of the extracted peak from the time-frequency analysis.
	(c) Speed of sound extracted from the dipole frequency as as function of blueshift, $\Delta E$,  for different excitation powers above the condensation threshold, $P_{\rm{th}}\approx10.5$~mW.  The black dashed line ($c = c_B/3$) and the blue dash-dotted line ($c = c_0 + c_1\Delta E$, where $c_0=0.4$~$\mu$m ps$^{-1}$ and $c_1=0.16$~$\mu$m ps$^{-1}$meV$^{-1}$) are guides to the eye corresponding to the expected square-root dependence and the observed linear dependence, respectively.
	}
\end{figure}

A remarkable advantage of a box trap is the dependence of the mode frequencies on the speed of sound, Eq.~\ref{eq:dispersion_law}, which for a 2D quantum gas is a function of its thermodynamic properties and the interaction strength \cite{Ota2018,DeRosi2015}. In order to analyze the frequencies of the two dominant oscillations, we perform a time-frequency analysis of the oscillation signals presented in Fig.~\ref{fig2}(c) and Fig.~\ref{fig3}(a) using a wavelet synchrosqueezed transform~\cite{Daubechies2016}. Fig.~\ref{fig4}(a) shows the two extracted down-chirped frequencies (see SM for more details). The low-frequency mode is extracted from $\langle x,y \rangle$ and is therefore due to the dipole oscillation, while the high-frequency component is extracted from $a,b$ and is due to the breathing oscillation.

The observed frequencies can be compared to those in a non-interacting, single-particle (sp) limit, where the oscillation is due to the linear superposition of the ground and excited states with either $m{=}1$ (dipole) or $m{=}0$ (breathing). These frequencies are plotted as dashed lines in Fig.~\ref{fig4}(a) (see SM for more details), where we estimate the trap radius $R$ from the running average of the measured Thomas-Fermi condensate width. Clearly, the observed frequencies are much higher than the non-interacting case, the ratio $\omega_B/\omega_D\approx2$, shown in Fig.~\ref{fig4}(b), is smaller than the non-interacting limit of $\omega^{\rm{sp}}_{B}/\omega^{\rm{sp}}_{D}\approx 2.7$, and the frequency chirp is reversed. Thus, the observed oscillation is collective in nature, in contrast to previously observed non-equilibrium motion of polaritons in a ring~\cite{Mukherjee2019}.

The breathing mode is a compressional mode of the 2D quantum gas, sensitive to the equation of state~\cite{DeRosi2015}.
For a 2D weakly interacting Bose gas in an elliptical box trap with the aspect ratio $a/b\approx 1.2$, one expects $\omega_B/\omega_D \approx 2$ (see SM). The ratio $\omega_B/\omega_D \approx 2.0$ observed in the polariton condensate [Fig.~\ref{fig4}(b)] therefore suggests that it behaves as a weakly interacting 2D Bose gas in a box trap.

The dipole mode can be used to measure the speed of sound using the dispersion law, Eq.~\ref{eq:dispersion_law}, and assuming that the condensate is in a quasi-steady state. This is a reasonable assumption because the condensate decays slowly, so that there are $>$10 oscillations per lifetime. At zero temperature and with negligible quantum depletion, the speed of sound should be equal to the Bogoliubov sound $c_B = \sqrt{gn/m}$, where $\mu=gn$ is the condensate interaction energy, $g$ is the polariton-polariton interaction strength, $n$ is the polariton density, and $m$ is the effective mass. The interaction energy in our experiment can be inferred from the instantaneous blueshift $\Delta E$ of the condensate energy [Fig.~\ref{fig1}(b)], provided that it arises only due to the polariton-polariton interaction~\cite{Estrecho2019}.

Fig.~\ref{fig4}(c) presents the measured speed of sound, $c$, as a function of the blueshift, $\Delta E$, for different excitation powers, where the blueshift is measured with respect to the polariton energy at $k=0$ in the low-density limit (see SM). As expected, the speed of sound decreases with the diminishing interaction energy (i.e., with time) and its general behaviour is independent of the excitation power, which only determines the initial polariton density and blueshift.
The results for different trap sizes and effective masses (see SM) show that, at early times, $c$ is independent of the trap size and decreases with increasing effective mass. At large blueshifts (or early times), $c(\Delta E)$ follows the predicted square-root law but with $c\approx c_B/3$. Furthermore, when $\Delta E\lesssim 1$~meV, this dependence deviates from the square-root law and becomes linear $c\propto \Delta E$.

The discrepancy between the measured and predicted speed of sound at large blueshifts can be attributed to the presence of the reservoir. In contrast to our previous work~\cite{Estrecho2019}, here the reservoir is not fully depleted, as evidenced by the slow decay of the condensate PL (see SM).
Therefore the polariton--reservoir interaction contributes to the measured blueshift, i.e. $\Delta E=g(n+|X|^{-2}n_R)$, where $n_R$ is the reservoir density, and $|X|^2$ is the excitonic Hopfield coefficient. Consequently, $c_B=\sqrt{gn/m} < \sqrt{\Delta E /m}$. Given the value for the interaction strength at near-zero detuning ($|X|^{2}=1/2$) \cite{Estrecho2019,Pieczarka2020}, we can estimate the ratio of densities to be $n_R/n \approx 4$ at early times and $n_R/n \approx 1.5$ at later times (see SM). The latter value agrees with previous measurements under off-resonant~\cite{Pieczarka2020} and resonant~\cite{Stepanov2019} CW excitation conditions. Therefore, the observed deviation from the square-root scaling at later times could indicate that both the reservoir and condensate densities approach a stationary state.

Although the oscillation frequencies $\omega_{B,D}$ and their ratio are in good agreement with the conservative theory, the damping of the excitations is the consequence of driven-dissipative and finite-temperature effects. The damping rates estimated from the data in Figs.~\ref{fig2}(d) and ~\ref{fig3}(a) (see SM) are $\gamma_D \sim 1$~ns$^{-1}$ for the dipole mode and $\gamma_B \sim 2$~ns$^{-1}$ for the breathing mode, resulting in a $Q$-factor $Q=\omega/\gamma\sim60$. The damping can be due to different mechanisms. Decay and driven thermalization of polaritons leads to damping of the excitations ~\cite{Keeling2004,Wouters2007,Byrnes2012,Smirnov2014} with the rate determined by the polariton lifetime and the stimulated scattering rate from the reservoir to the condensate (see SM). While the latter is not known, a reasonable estimate (see SM) leads to the damping rates on the order of~ns$^{-1}$, which are similar to the rates observed in the experiment. Similar rates can also arise from excitations of the excitonic reservoir \cite{Keeling2004,Wouters2007,Byrnes2012,Smirnov2014} (see SM). Furthermore, the excitations can be damped by scattering with lattice phonons~\cite{Wouters2012}, which results in an effective energy relaxation for polaritons~\cite{Wertz2012, Bobrovska2014, Estrecho2019}. A similar relaxation process has been shown to damp condensate oscillations in conservative (cold atom) systems~\cite{Choi1998}. However, these effects cannot fully account for the observed dependence of the damping rates on the momentum. Interaction with uncondensed, thermal polaritons, which are observable in high quality samples and have an effective temperature of $T\sim$10~K \cite{Pieczarka2020,Sun2017}, such that $k_B T\sim gn \gg \hbar \omega_{B,D}$, can lead to momentum-dependent Landau damping~\cite{Giorgini1998,Ota2018b, Cappellaro2018}. The measured Q-factors of the two modes are similar to those reported in previous studies of Landau damping of collective oscillations in a conservative 2D Bose gas in the collisionless regime~\cite{Ville2018,Ota2018b, Cappellaro2018}. Indeed, our condensate is in the interaction-driven collisionless rather than the hydrodynamic regime~\cite{Ota2018b} because the effective collision frequency~\cite{Petrov2001} $\Omega\sim 0.1-1$~ns$^{-1}$ is much smaller than $\omega_{B,D}$ (see SM).

{\em Conclusion.---}We have observed collective oscillations of a polariton condensate in a box trap. The oscillations
are dominated by the dipole and breathing modes, with the ratio of frequencies well described by a model of 2D weakly interacting bosons. The speed of sound determined from the dipole frequency is lower than the Bogoliubov sound, assuming the condensate blueshift is only due to polariton-polariton interaction. This discrepancy points to the significant influence of the reservoir. 

Our future work will focus on selective excitation of collective modes by pulsed perturbation of a steady-state condensate. This will allow us to relate the dispersion of excitations~\cite{Pieczarka2020} to the measured speed of sound, determine the momentum dependence of the damping rates, and identify the dominant damping mechanism.

Our study paves the way for further investigations of the collective excitations of polariton condensates, which are essential for better understanding of the driven-dissipative and finite temperature effects ~\cite{Wouters2010, Cappellaro2018, Ota2018b} on the superfluidity of 2D non-equilibrium systems. Precise measurements of the breathing mode frequency can lead to experiments on quantum corrections beyond the mean-field approximation ~\cite{Peppler2018, Holten2018}, and enable tests of a crossover between the quantum phases of polariton systems.

\bibliographystyle{apsrev4-2}

\end{document}



\title{Low-energy collective oscillations and Bogoliubov sound in an exciton-polariton condensate \\ (Supplemental Material)}

\author{E.~\surname{Estrecho}}
\author{M.~\surname{Pieczarka}}
\author{M.~\surname{Wurdack}}
\author{M.~\surname{Steger}}
\author{K.~\surname{West}}
\author{L.~N.~\surname{Pfeiffer}}
\author{D.~W.~\surname{Snoke}}   
\author{A.~G.~\surname{Truscott}}
\author{E.~A.~\surname{Ostrovskaya}}

\maketitle
\begin{figure}[h]
\includegraphics[width=\columnwidth]{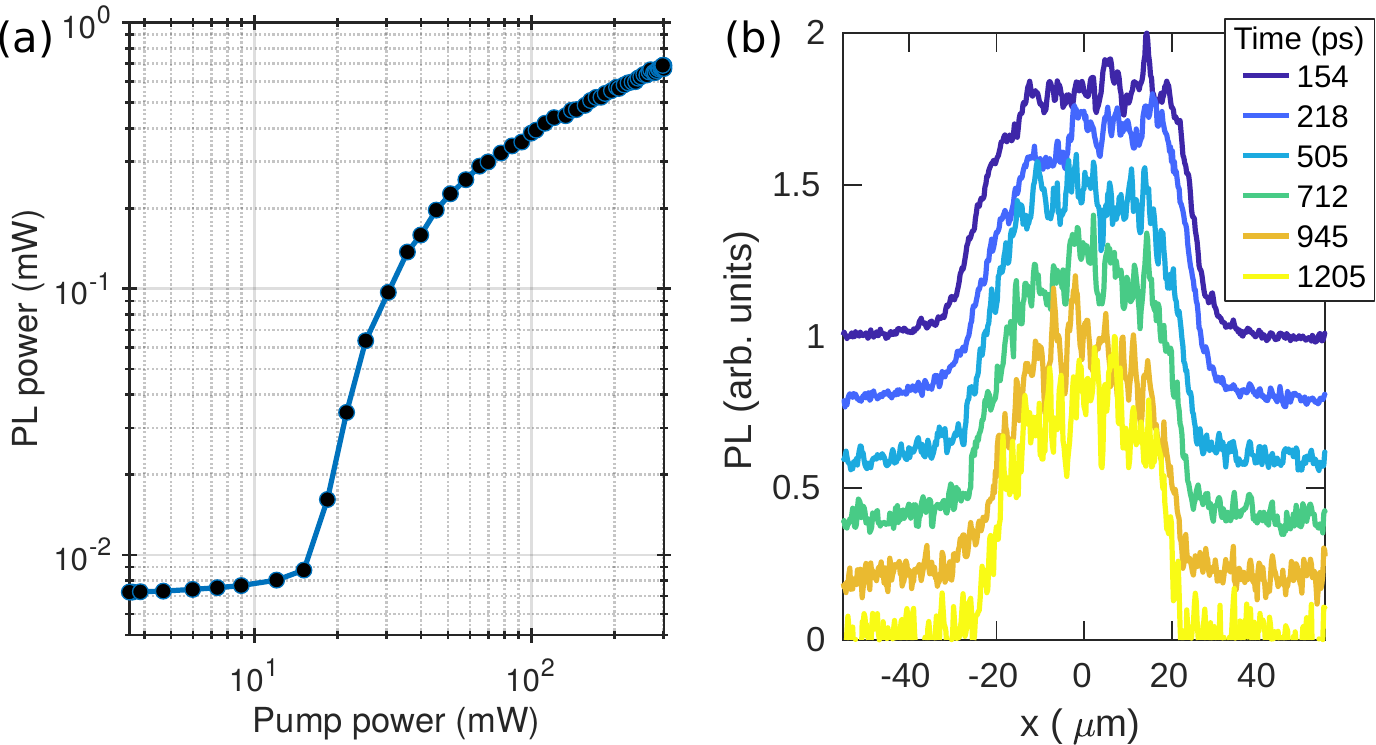}
\caption{\label{fig:power} (a) PL as a function of pump power. The condensation threshold occurs at $P_{th}\approx 15$~mW. (b) Condensate spatial profile at different times corresponding to Fig.~1(d) of the main text.}
\end{figure}

\textbf{Experimental details:}
The experimental setup is similar to our previous work~\cite{Estrecho2019}, where the excitation beam is shaped into an annular distribution using an axicon lens~\cite{Pieczarka2019prb} with two notable modifications, namely time-resolved measurements and higher-energy excitation ($150$~meV above the polariton resonance). We perform time-resolved imaging using a streak camera (Optronis SC-10) synced to an 80 MHz pulsed laser (Coherent Chameleon Ultra II) with a nominal 140 fs pulse duration.
An example input-output power characteristics of the studied polariton system is presented in Fig.~\ref{fig:power}(a).

Time-resolved tomography is achieved by translating the imaging lens in front of the monochromator coupled to the streak camera. This translation is performed to record the time-resolved real-space (RS) $n_r(x,y,t)$ and $k$-space (KS) $n_k(x,y,t)$ distributions. The supplementary movie file \texttt{Supplementary\_Movie.avi} shows an example of the normalized RS and KS distributions of polaritons. The sharp edges of the condensate are also evident in the spatial profiles $n_r(x,y{=}0,t)$ at different times $t$ as shown in Fig.~\ref{fig:power}(b).

\begin{figure}[h]
\includegraphics[width=\columnwidth]{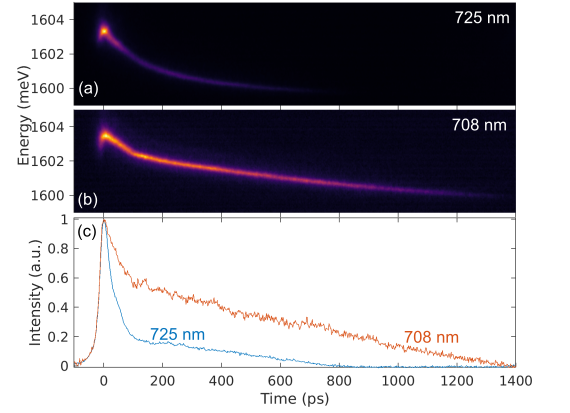}
\caption{\label{fig:pumpwav} Energy and time-resolved PL at the center of the condensate created by optical excitation at (a) 725 nm and (b) 708 nm with the power adjusted to achieve similar maximum blueshift. (c) Time-resolved PL under the two excitation conditions. The zero time is set to maximum intensity.}
\end{figure}

The excitation wavelength directly affects the decay time of the condensate photoluminescence (PL), as shown in Fig.~\ref{fig:pumpwav} for the excitation wavelengths of 725~nm and 708~nm at approximately equal maximum blueshift. Both exhibit two decay times which can be characterized by a two-reservoir model~\cite{Anton2013}. The fast decay is due to strong stimulated scattering from the reservoir to the condensate, which quickly depletes the ``active" reservoir that directly feeds the condensate. The slow decay is due to the continuous replenishment of the condensate by the slowly decaying ``inactive" reservoir via the active reservoir. The relaxation from the inactive to active reservoir depends on the excitation energy as demonstrated in time-resolved studies of exciton PL in GaAs-based quantum wells~\cite{blom1993,bacher1995,kovac1996}. The higher-energy excitation at 708 nm clearly shows a slower decay retaining 20\% of the signal up to 1~ns after the peak intensity compared to 100~ps for the excitation at 725 nm. In our experiments, we used an excitation wavelength around 708 nm to achieve long decay times while maintaining substantial laser power.

\begin{figure}[h]
\includegraphics[width=\columnwidth]{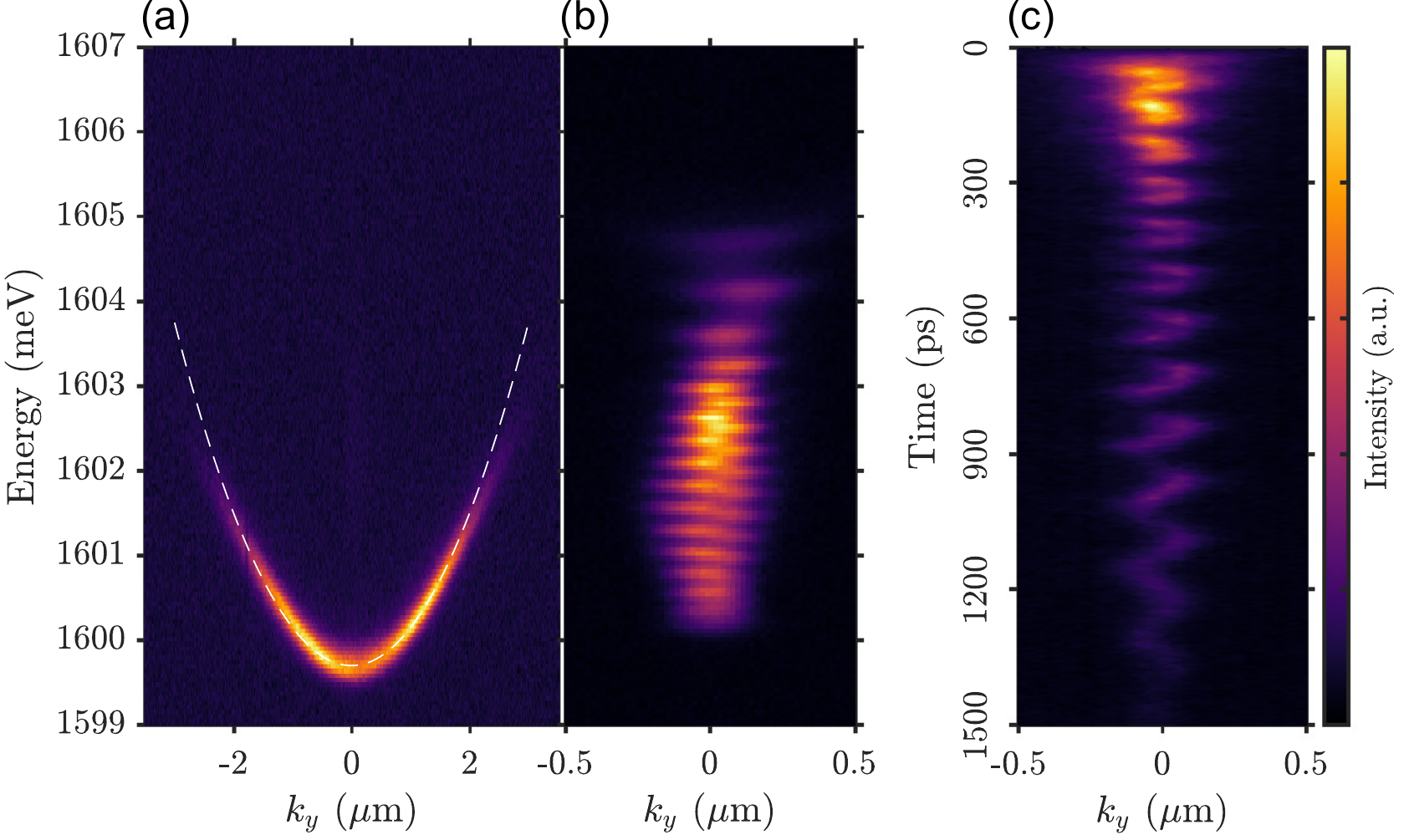}
\caption{\label{fig:KS} (a) Lower-polariton dispersion measured below the condensation threshold. The dashed line is the quadratic fit used to extract the effective mass (see text).
 (b) Energy-resolved and (c) time-resolved KS distribution of the polaritons shown in Fig.~1(c,d) of the main text.}
\end{figure}

\textbf{Effective mass fit and oscilllations in KS:}
Fig.~\ref{fig:KS}(a) shows a typical lower-polariton dispersion measured at a very low density below the condensation threshold. Using a parabolic fit to the dispersion with the formula: $E(k) = E_0 + \hbar^2 k^2 / 2m$, where $m$ is the polariton effective mass, we extract the value of $m=8.5\times 10^{-5}$ of the electron rest mass.

The energy-resolved and time-resolved KS distributions of the polaritons shown in Fig.~1(c,d) of the main text are presented in Fig.~\ref{fig:KS}(b,c).

\textbf{Ellipse fitting:}
To account for the asymmetry of the trap and the condensate, we fit the condensate shape to an ellipse. We find the edge of the condensate using the $1/e$ of the density maximum then fit the contour to a tilted ellipse centered at $x_0,y_0$ with the semi-axes lengths $a,b$.

 \textbf{Determination of the damping rates:}
Estimating the damping rates from the experiment is a nontrivial task due to the non-stationary behaviour of the oscillations. The oscillation amplitude increases at early times then decreases at later times. This is further complicated by the monotonic decrease of the oscillation frequency with time, which can result in a time-dependent damping rate. However, we can still, albeit roughly, approximate the decay rate in a short time range after the peak amplitude is attained, which occurs at around $500$ ps.

Looking at the oscillation in Fig. 2(d), we can see that the dipole oscillation amplitude decreases to 50\% around $1200$ ps. Assuming a constant exponential decay rate in this time range, this results in a decay constant of $1$ ns. For the breathing mode oscillation shown in Fig. 3(a), the breathing mode amplitude decreases below the noise level at around $1000$ ps. With the noise level approximately 25\% of the peak amplitude, we can roughly estimate the decay constant of $0.5$ ns.

\textbf{Polariton excitation spectrum:}
The excitation spectrum of polariton condensates at small wavevectors differs from the equilibrium Bogoliubov spectrum due the non-equilibrium nature of the system~\cite{Wouters2007}. Although in our experiment the condensate is trapped, the high-density Thomas-Fermi regime in the box trap allows us to employ the analytical theory for a spatially homogeneous condensate in the steady state regime. The (complex) excitation frequencies are given by the solutions of the cubic equation~\cite{Smirnov2014}:
\begin{equation}
\omega^3+i\gamma_R(1+\alpha)\epsilon^2 - [\omega_0^2+\alpha\gamma\gamma_R]\omega = f(k),
\end{equation}
where $f(k)=i\gamma_R(1+\alpha)\omega_0^2 -i\gamma k^2 g_R/g$, and $\omega_0^2=\frac{k^2}{2m}\left(\frac{\hbar^2k^2}{2m} + 2gn\right)$ is the Bogoliubov dispersion relation. The parameters are the polariton (reservoir) decay rate $\gamma$ ($\gamma_R$), the ratio of polariton--reservoir $g_R$ and polariton--polariton $g$ interaction strengths $g_R/g = |X|^{-2}$ that depends only on the excitonic Hopfield coefficient $|X|^2$, and the relative pump power $\alpha = P/P_{th} - 1$, where $P_{th}$ is the pump power at the condensation threshold.

\begin{figure}[h]
\includegraphics[width=1\columnwidth]{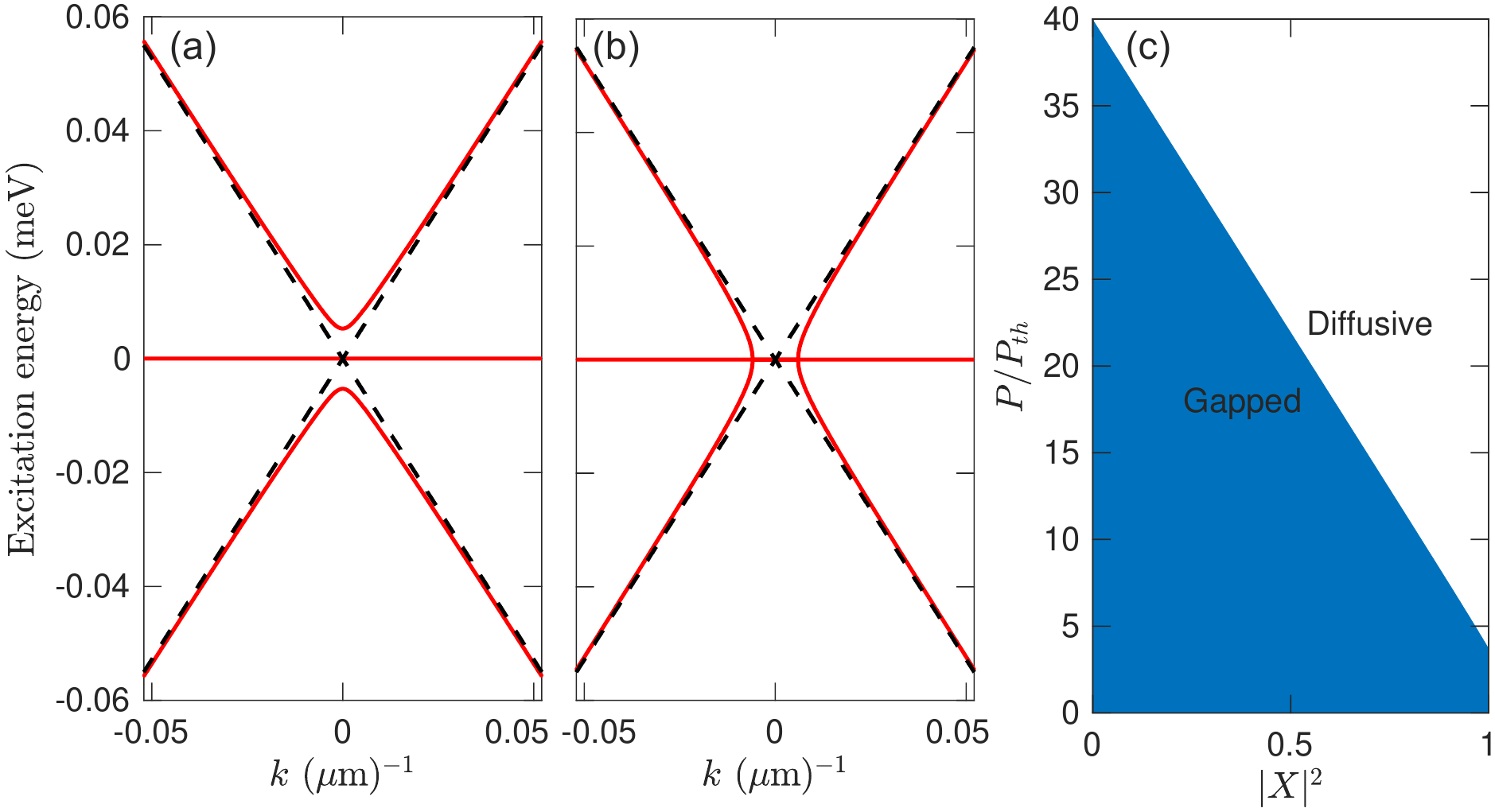}
\caption{\label{fig:excitation} The low-$k$ excitation spectra, $\epsilon(k)=\hbar\textrm{Re}(\omega)$, of a homogeneous polariton condensate in the (a) gapped ($P/P_{th}=10$) and (b) diffusive regimes ($P/P_{th}=40$). Dashed lines correspond to the Bogoliubov excitation spectra. (c) Domains where the system excitations are ``gapped" or ``diffusive". The parameters are: $m=8.5\times 10^{-5}$, $gn=1$~meV, $\gamma=1/100$~ps$^{-1}$, and $\gamma_R=1/1000$~ps$^{-1}$.}
\end{figure}

The dissipation and interaction with the reservoir modify the real part of the excitation frequency (dispersion) near $k=0$, which can be ``gapped'' or ``diffusive''~\cite{Byrnes2012,Smirnov2014} depending on the pump power relative to the condensation threshold, as shown in Fig.~\ref{fig:excitation}(a,b). The boundary between the two regimes, given by
$$
\frac{(P_{gap}/P_{th})^2}{(P_{gap}/P_{th})-1} = 4\frac{\gamma}{\gamma_R},
$$
is plotted in Fig.~\ref{fig:excitation}(c) for the set of parameters relevant to the sample used in this work. At lower pump powers, i.e. in the gapped regime, the excitations have an effective mass, but are nondispersive at higher pump powers, i.e. in the diffusive regime. The wavevector range $0<|k|<k_s$, where this deviation from the Bogoliubov spectrum occurs, depends on the polariton decay rate $\gamma$. Larger $\gamma$ or shorter lifetime leads to larger $k_s$. For the sample used in this work, the polariton decay rate is $\gamma<1/100$~ps$^{-1}$ and the gapped or diffusive regions occurs only at $|k|<k_s\approx 0.01~\mu$m$^{-1}$ [see Fig.~\ref{fig:excitation}(a,b)]. This is an order of magnitude smaller than the wavevectors probed in this work, which is $\sim 0.1~\mu$m$^{-1}$ for the dipole mode. Outside this range, the excitation spectrum can be well approximated by the Bogoliubov dispersion [dashed lines in Figure~\ref{fig:excitation}(a,b)]. We would need a very large condensate $R>100~\mu$m to access the values of momenta in the non-sonic range, $k<k_s$, in our experiment. Hence, in our experiment, the oscillations are well in the sonic regime and it is sufficient to use the Bogoliubov theory to analyze the oscillation frequencies.

Importantly, the damping rates characterised by the imaginary part of the excitation frequencies $\Gamma=\textrm{Im}(\omega)$ are non-zero for both the condensate and reservoir excitations for the whole range of wavevectors, both in the gapped and the diffusive regime. For $k>k_s$ and the relevant experimental parameters, the decay rates for the condensate oscillations and aperiodic reservoir excitations are almost constant, and can be estimated as $\Gamma\approx \hbar s_R \gamma/(2g)$ and $\Gamma_R\approx \gamma_R-2\gamma+ns_R$, respecively, where $s_R$ is the effective stimulated scattering rate from the reservoir to the condensate. While the latter value is not known, a realistic estimate of $s_R\sim 10^{-5}-10^{-4}$ $\mu$m$^2$/ps \cite{latticePRB} yields the decay rates similar to the measured values of $\Gamma_{B,D}$. However, the very weak $k$-dependence of the decay rates predicted by this theory is contrary to the factor of two difference between $\Gamma_B$ and $\Gamma_D$ observed in the experiment. Therefore, other momentum-dependent damping mechanisms should be investigated.

\textbf{Landau damping and the collisionless regime:}
Thermal excitations can play an important role in the condensate dynamics resulting in the Landau damping of oscillations. The damping arises from the absorption of quanta of oscillations by thermal excitations. Although the condensate is deeply degenerate, with densities up to four orders of magnitude higher than the threshold density~\cite{Estrecho2019}, the thermal energy (assuming $T\sim 10~K$, which is the lattice temperature) is comparable to the interaction energy, i.e. $\tau = k_B T/gn\sim 1$. Moreover, $k_B T$ is an order of magnitude higher than the oscillation energy $\hbar\omega\sim0.1$~meV. Hence, the thermal excitations can play a dominant role in the observed oscillations. Following Ref.~\cite{Chung2009}, we can estimate the damping rate in the hydrodynamic regime as $\gamma \approx 0.02 \tilde{g} \tau \omega $, where the dimensionless interaction parameter is $\tilde{g}=mg/\hbar\sim10^{-3}$. For the system considered here, we use $g\approx 2\mu$eV$\cdot\mu$m$^2$ as the polariton-polariton interaction strength per quantum well~\cite{Estrecho2019}. This results in a damping rate $\gamma\sim 10^{-3}$~ns$^{-1}$, which is 3 orders of magnitude lower than what was observed.

However, it is important to point out that the observed superfluid dynamics is in the collisionless regime~\cite{Ota2018b}. The interparticle collision frequency~\cite{Petrov2001} is $\Omega=\hbar{\tilde{g}}^2n/m\sim 0.1-1$~ns$^{-1}$ for polariton densities per quantum well of $n\sim10^2-10^3~\mu$m$^{-1}$ or $gn\sim 0.1-1$~meV.
Since the oscillation frequency $\omega$ is much greater than the collisional rate, i.e. $\omega \gg \Omega$, local equilibrium is not ensured so the system is in the collisionless regime instead of the hydrodynamic regime. The observed Q-factors are in good agreement with the damping observed in recent experimental and theoretical works on conservative 2D Bose gas~\cite{Ville2018, Cappellaro2018, Ota2018b}.

\begin{figure}[h]
\includegraphics[width=0.6\columnwidth]{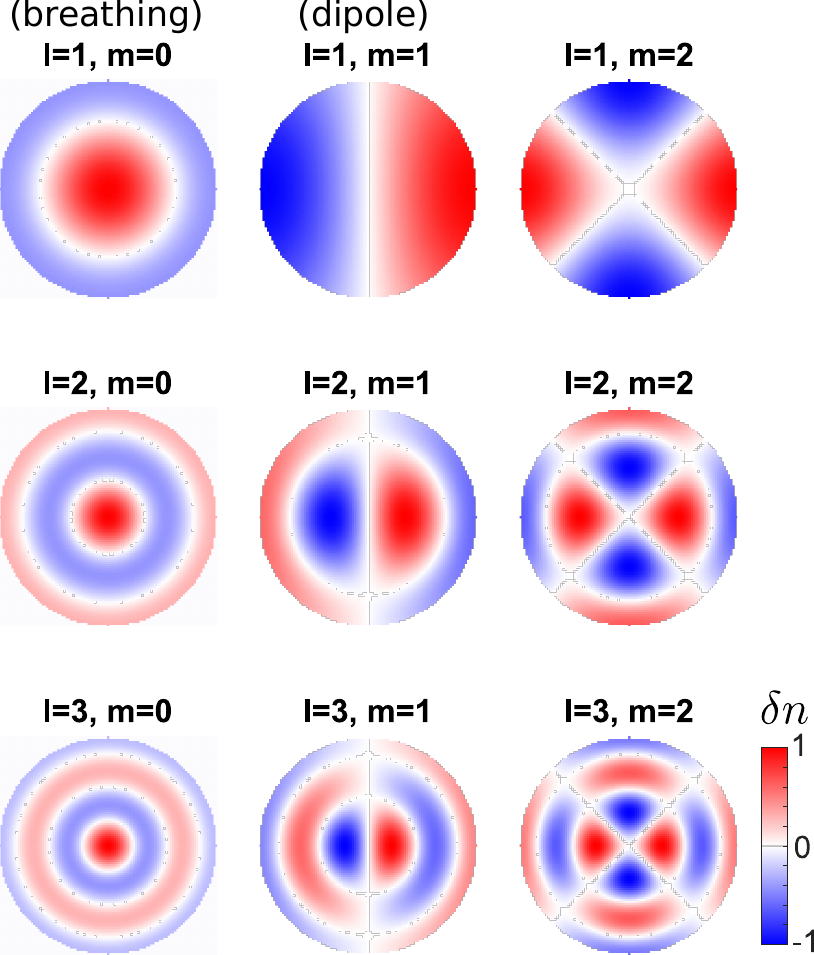}
\caption{\label{fig:collective_modes} Spatial profiles of the low-lying excitation modes $\delta n_{l,m}$ of a weakly-interacting condensate in a circular box trap. Modes with negative $m$ have the same profiles but rotate in the opposite direction.}
\end{figure}

\textbf{Excitation modes of a condensate in a circular box trap:}
Following the main text, the excitation modes of the condensate in a circular box trap of radius $R$ have the approximate form
$$
\delta n_{l,m} \propto J_m\left( \frac{q_{l,m}r}{R}\right) e^{im\phi}
$$
with the angular frequencies
$$
\omega_{l,m} = c \frac{q_{l,m}}{R},
$$
where $c$ is the speed of sound, $J_m$ is the Bessel function of the first kind, and $q_{l,m}$ is the $l$-th non-zero root of the derivative of the Bessel function $J'_m$.
In the linear approximation, i.e. when the excitation modes are decoupled from each other, the spatio-temporal evolution of the condensate distribution is described by the real part of the expression:
\begin{equation}
\label{eq:sum_oscillation}
n(x,y,t) \approx n_0(x,y) + \sum_{l,m} c_{l,m} \delta n_{l,m}(x,y) e^{i\omega_{l,m}t},
\end{equation}
where  $c_{l,m}$ is the contribution of the $\delta n_{l,m}$ mode and $n_0(x,y)$ is the ground state density distribution. The spatial profiles of the low-energy modes of the trap are shown in Fig.~\ref{fig:collective_modes}.
Note that the modes with $m \neq 0$ rotate with time resulting in a centre-of-mass motion.

The dipole mode ($l{=}1,m{=}1$) has the lowest frequency with $\omega_D = 1.8412 {c}/{R}$. The breathing mode ($l{=}1,m{=}0$) has the frequency $\omega_B = 2.081 \omega_D$ while the quadrupole mode ($l{=}1,m{=}2$) has the frequency $\omega_Q = 1.659 \omega_D$. Note that the latter was not observed in the experiment.

\begin{figure}[h]
	\includegraphics[width=0.6\columnwidth]{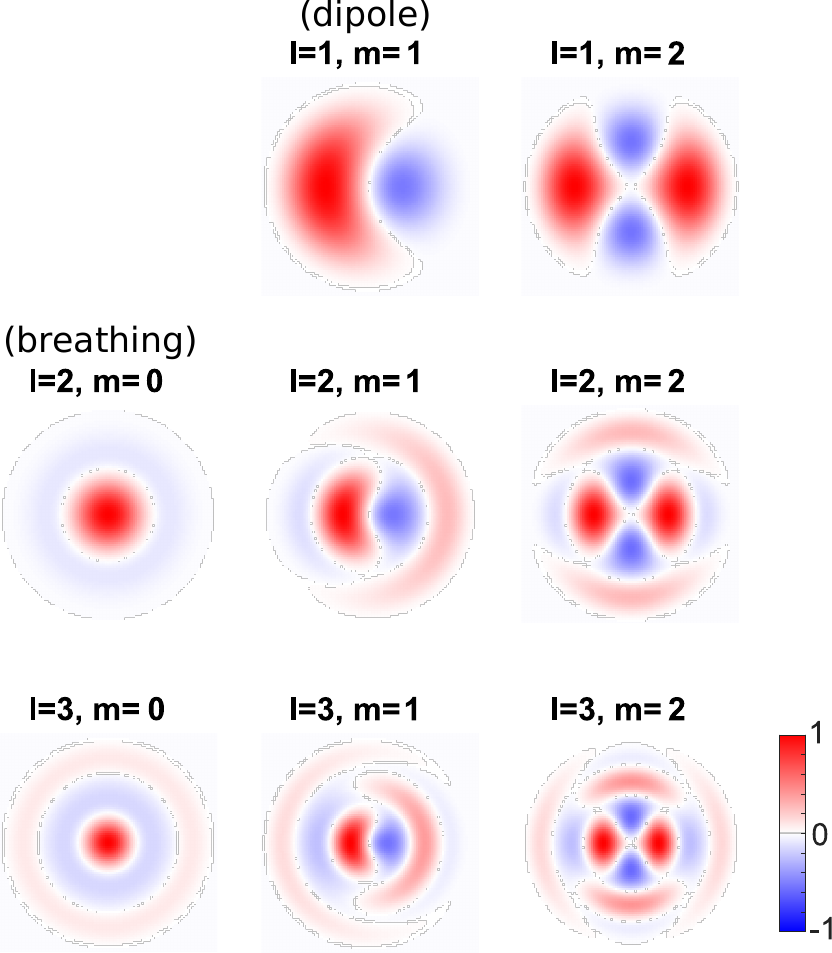}
	\caption{\label{fig:singleparticle_modes} Density profiles of the non-interacting oscillation modes of a particle in a circular box trap resulting from the superposition of the ground state $\delta n^{(sp)}_{1,0}$ and the excited state $\delta n^{(sp)}_{l,m}$. For better visualization, the ground state density $|\delta n^{(sp)}_{1,0}|^2$ is subtracted and the time $t$ is chosen to maximize the deviation from the ground state.}
\end{figure}

\textbf{Oscillation modes in the non-interacting case:}
The oscillations of a non-interacting quantum gas can arise from the superposition of single-particle states of the trap $\delta n^{(sp)}_{l,m}$, which can be written as
$$
\delta n^{\rm{sp}}_{l,m} \propto J_m\left( \frac{q^{\rm{sp}}_{l,m}r}{R}\right) e^{im\phi}
$$
where $q^{\rm{sp}}_{l,m}$ is the $l$-th root of the Bessel function $J_m$ and the eigenenergies are
$$
E_{l,m} = \frac{\hbar^2}{2mR^2} \left( q^{\rm{sp}}_{l,m} \right)^2.
$$
The low-energy oscillation modes are superpositions of the ground state $\delta n^{(sp)}_{1,0}$ and the excited state $\delta n^{(sp)}_{l,m}$, and the time-dependent density distribution is
$$
n_{l,m}(r,\phi,t) \propto \left| \delta n^{\rm{sp}}_{1,0}e^{iE_{1,0}t/\hbar} + \alpha \delta n^{\rm{sp}}_{l,m}e^{iE_{l,m}t/\hbar}\right|^2
$$
where $|\alpha|^2$ is the weight of the excited state. These modes are shown in Fig.~\ref{fig:singleparticle_modes} with the ground state density subtracted, i.e. $n(r,\phi,t) - |\delta n^{\rm{sp}}_{1,0}|^2 $, for better visualization of the modulations on top of the ground state. The characteristic angular frequencies of these oscillations can be written as
$$
\omega^{\rm{sp}}_{l,m} = \frac{E_{1,0}-E_{l,m}}{\hbar} = \frac{\hbar}{2mR^2} \left( [q^{\rm{sp}}_{l,m}]^2 - [q^{\rm{sp}}_{1,0}]^2 \right),
$$
which is simply the frequency difference of the two states.

The lowest-frequency mode is the dipole mode ($l=1, m=1$) with the frequency $\omega_D^{(sp)} = 8.899\hbar/(2mR^2)$. The lowest breathing mode ($l=2, m=0$) has the frequency $\omega_B^{(sp)} = 2.774 \omega_D$, while the quadrupole mode ($l=1, m=2$) has the frequency $\omega_Q^{(sp)} = 2.314\omega_D$.
Note that the ratio of the frequencies of the modes is significantly different for the weakly-interacting and the non-interacting case, hence it can be used to differentiate between the collective and single-particle oscillations.

\textbf{Gross--Pitaevskii simulations:}
To confirm that the centre-of-mass and width oscillations observed in the experiment arise from dipole and breathing modes, we perform numerical simulations of the conservative Gross--Pitaevskii equation given by
$$
i\hbar \frac{\partial \Psi}{\partial t} = \left[ -\frac{\hbar^2 \nabla^2}{2m}  + V + g|\Psi|^2 \right] \Psi
$$
for the condensate wavefunction $\Psi(x,y,t)$ given an external potential $V(x,y,t)$ by implementing the MATLAB GPELab Toolbox of Refs.~\cite{antoine2014gpelab,antoine2015gpelab}.
This model does not take into account dissipation and only works in the zero-temperature limit where the thermal depletion of the condensate is negligible.

There are two steps in the simulation. First, the ground state is calculated for a given potential and the number of particles. In all cases shown here the potential is approximated by an inverted super-Gaussian function, i.e.:
$$
V(x,y) = V_0 \left(1-\exp{\left[-\left(\frac{x^2}{a^2} + \frac{y^2}{b^2}\right)^\eta \right]} \right),
$$
where $V_0$ is the depth of the potential, $a,b$ are the semi-axes of its cross-section, and $\eta$ is the shape parameter, with $\eta \gg 1$ for a box-like and $\eta=1$ for a harmonic-like potential.
Secondly, a time-dependent potential in the form of a short pulse within the first 10 ps of the simulation is used to excite the oscillation modes.
To avoid exciting a myriad of modes, we use a weak excitation potential with a spatial profile matching the target mode.
For the dipole mode, the excitation is simply a tilt resulting in a sinusoidal motion of the center-of-mass (COM) of the condensate.
A spatial Gaussian pulse concentric with the trap excites the breathing mode resulting in a sinusoidal motion of the width of the condensate while ensuring no COM motion.

\begin{figure}[h]
	\includegraphics[width=\columnwidth]{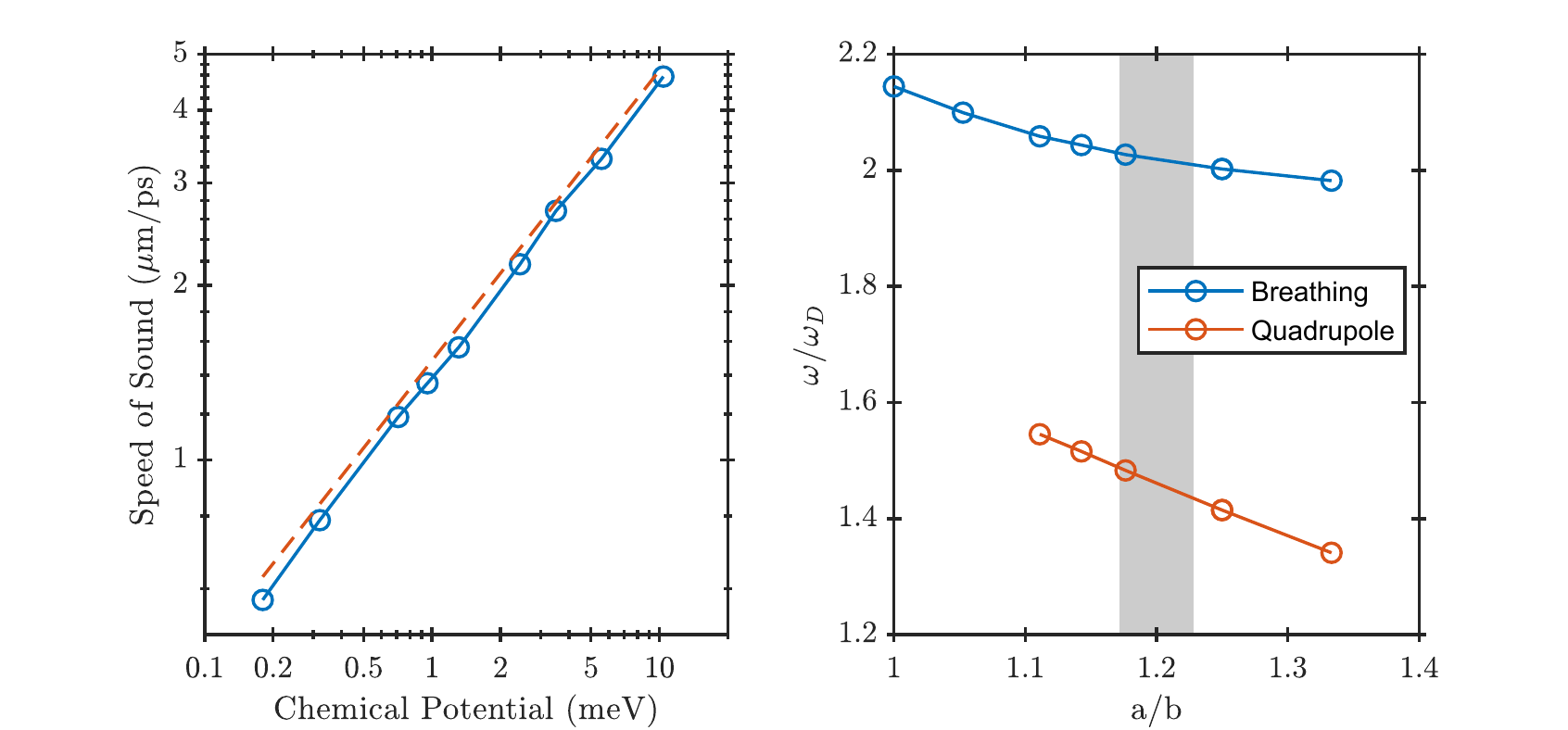}
	\caption{\label{fig:simulation} (a) Speed of sound (open circles) extracted from simulations. Dashed line is the Bogoliubov sound $c_B=\sqrt{gn/m}$.
	(b) Frequencies extracted from the width oscillation $\omega$ in units of extracted dipole frequency $\omega_D$ of an elliptial trap. Shaded region is the aspect ratio of the condensate in the experiments.}
\end{figure}

The first set of simulations is aimed at confirming the validity of the extraction of the speed of sound from the dipole oscillation. We use a round trap with the radius $a=b=20~\mu$m and $\eta=10$, and vary the number of particles in the trap. The chemical potential $\mu = gn$, the radius of the condensate $R$ measured as the half-width at $1/e$ of maximum, and the frequency $\omega_D$ of the resulting COM motion are then extracted from the simulation. Fig.~\ref{fig:simulation}(a) confirms that we can reliably extract the speed of sound using the formula $c = \omega_D R/1.8412$, which we compared to the Bogoliubov sound $c_B = \sqrt{gn/m}$ for the parameters used in the simulation.

The second set of simulations is used to examine the deviation of the ratio of the breathing and dipole frequencies $\omega_B/\omega_D$ from the theoretical approximation used in the main text and how this value changes with the aspect ratio $a/b$.
To this end, we fix $\eta=10$, $a=20~\mu$m and the number of particles, tune the short-axis length $b$, separately excite dipole and breathing modes, and extract the mode frequencies.
The tilt excitation is oriented along the short axis of the trap so that we only excite the dipole mode along that direction.
The results are presented in Fig.~\ref{fig:simulation}(b) showing two modes extracted from the width oscillations.
We attribute the higher frequency to the breathing mode and the lower frequency to the quadrupole mode since their values approach the theoretical estimates $\omega_B/\omega_D \approx 2.08$ and $\omega_Q/\omega_D \approx 1.67$, respectively, at $a=b$.
Note that the signal from the quadrupole mode vanishes when $a/b=1$, which explains why it is not visible in the experiment.
More importantly, when the trap is circular, the ratio is $\omega_B/\omega_D\approx 2.14$, which is higher than the theoretical estimate. This can be due to the sharp edges of the condensate, which are neglected by our theory.
However, the frequency ratio approaches $\omega_B/\omega_D\approx 2.0$ for the aspect ratio of $a/b\approx1.2$, in good agreement with the experimental results.

\textbf{Principal component analysis (PCA):}
We employ a powerful model-free statistical analysis tool called principal component analysis (PCA)~\cite{Segal2010,Dubessy2014} to extract and identify the oscillation modes in the experiment. For each measurement with a certain set of experimental parameters, we record the time-resolved RS distribution (using tomography with a streak camera) of the microcavity emission. The acquired signal is a three-dimensional (3D) intensity array which we recast into a series of 2D arrays $I_t(x_i,y_j)$ where $I_t$ is the intensity at pixel $(x_i,y_j)$ at a time $t$. To reduce the noisy contributions from pixels outside the condensate emission, we select a small region of interest (ROI) that includes the entire condensate.

Given a time series of $N_t$ images $I_t(x_i,y_j)$, PCA finds principal components $\mathcal{P}_c(x_i,y_j)$ (which are also images) that are \textit{uncorrelated} with each other such that the original data can be written as a linear combination of these components, i.e.:
$
I_t(x_i,y_j) = \overline{I(x_i,y_j)} + \sum_{c=1}^{N_t} w_{c,t} \mathcal{P}_c(x_i,y_j),
$
where $w_{c,t}$ is the amplitude of the $c$-th principal component, and $\overline{I(x_i,y_j)}$ is the time-averaged signal.
In the RS representation, it simply means that the time-resolved RS distribution can be written as: 
\begin{equation}
\label{eq:PCA_RS}
n_r(x,y,t) = \overline{n_r(x,y)} + \sum_{c=1}^{N_t} w_{c}(t) \mathcal{P}_c (x,y).
\end{equation}
By comparison with Eq.~\ref{eq:sum_oscillation}, it is clear that the principal components $\mathcal{P}_c(x,y)$ correspond to the oscillation mode profiles $\delta n(x,y)$, and the time dependence is embedded in the coefficients, such that $w_{c}(t) \propto \cos(\omega_{l,m}t)$. For the mathematical details of the method and its application to the dynamics of atomic condensates, we refer the reader to Refs.~\cite{jolliffe2011principal,Dubessy2014} and references therein.

\begin{figure}[h]
	\includegraphics[width=0.7\columnwidth]{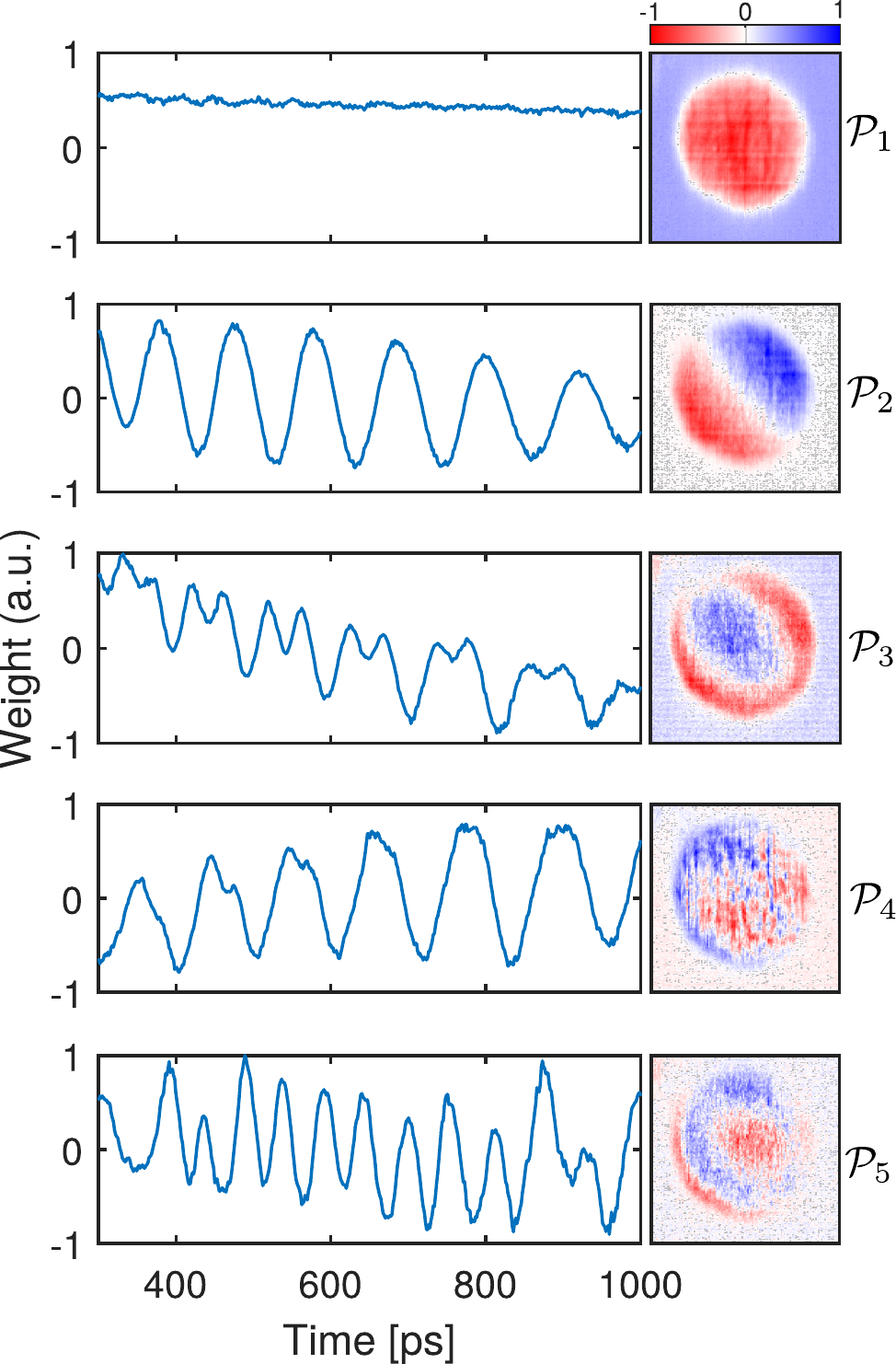}
	\caption{\label{fig:PCA} Principal components extracted from the PCA of the time-resolved RS distribution. (left) Amplitudes and (right) spatial distributions of the principal components PCs.}
\end{figure}

In this work, the actual numerical implementation of PCA is performed using a built-in MATLAB function~\cite{MATLAB2019a}. A plot of the amplitudes and the spatial distributions of the first 5 principal components (PCs), sorted by descending order of component variance, are shown in Fig.~\ref{fig:PCA} for a time range of $300$--$1000$~ps. The first component $\mathcal{P}_1$ corresponds to the ground state condensate, which is not oscillating but shows a decreasing amplitude.
The two relevant principal components are $\mathcal{P}_2$ and $\mathcal{P}_3$, which correspond to the dipole and breathing modes presented in the main text.
$\mathcal{P}_4$ is the second dipole mode orthogonal to $\mathcal{P}_2$, displaying a similar frequency, as expected.
$\mathcal{P}_5$ looks like a higher order mode, but its frequency is similar to that of $\mathcal{P}_3$.
The rest of the principal components have frequencies equal to the dipole or the breathing mode frequencies, suggesting that these modes are part or linear combinations of the two main modes.

We also performed PCA on different time ranges, but the extracted modes deviate strongly from theory without clear oscillatory behavior, especially when the early times of the condensate dynamics are included. This is due to the abrupt changes in the intensity and size of the condensate during those times. Hence, we limit the analysis to the time ranges that show clear oscillations that also agree with the oscillations observed in the COM and width of the condensate (see Fig.~3 of the main text).

\begin{figure}[th]
	\includegraphics[width=\columnwidth]{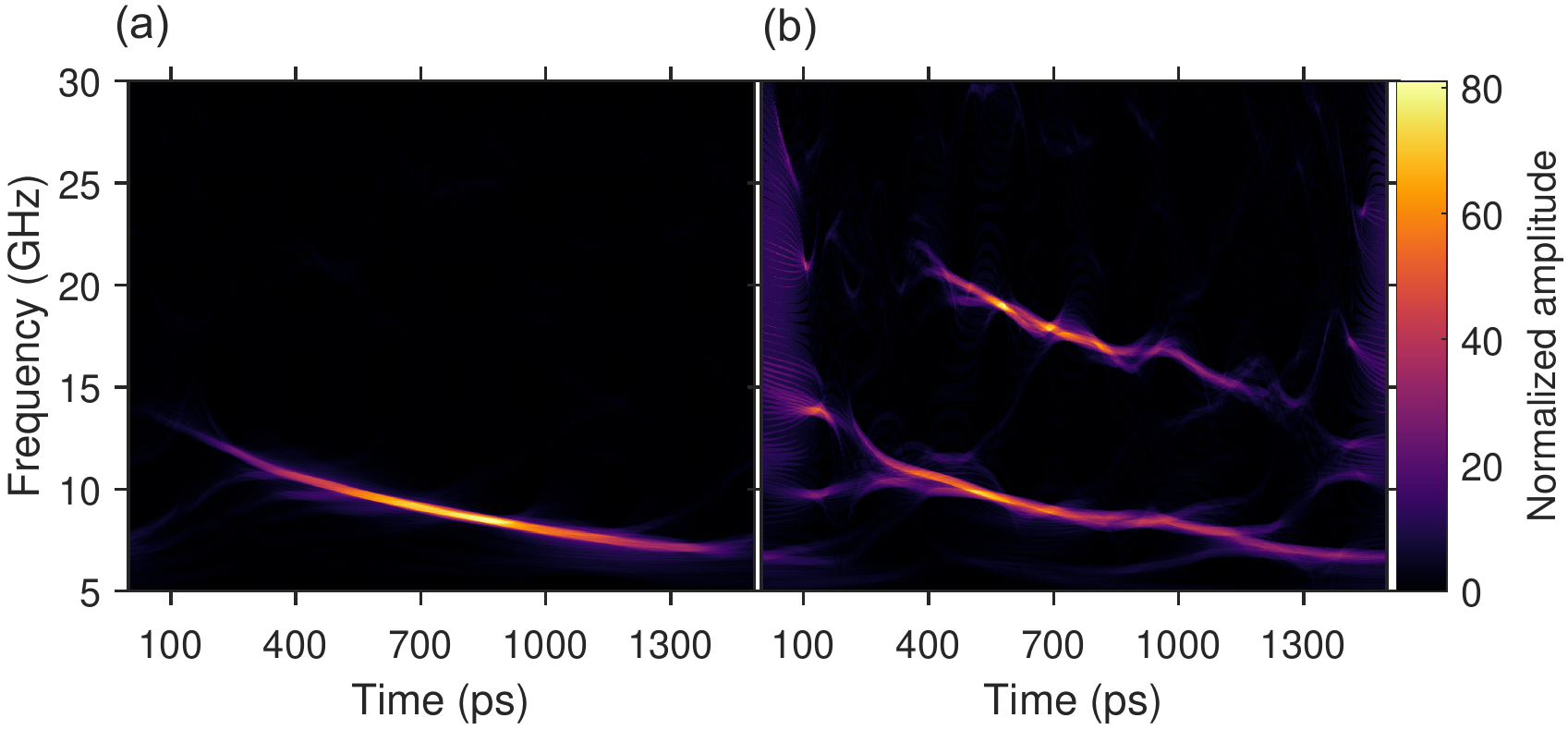}
	\caption{\label{fig:TF} Time-frequency representations of the oscillation data shown in Fig. 4(a) of the ellipse parameters (a) $x_0$ and (b) $b$.}
\end{figure}

\textbf{Time-frequency analysis:}
The oscillations measured in the experiment display time-dependent and dual frequency behavior. This makes traditional frequency analysis, like Fourier analysis or direct fitting of the signal, inefficient in finding the instantaneous frequency with a high precision. One can use the short-time Fourier transform, which uses small time windows, but it does not offer a good resolution for signals with few cycles, and is limited to quasi-stationary signals. To overcome these difficulties, we employed the synchrosqueezed continuous wavelet transform (SCWT)~\cite{Dubessy2014,Daubechies2016} which features a sharp time-frequency representation of the signal to enable both precise measurement of the instantaneous frequency and separation of different modes. The actual implementation of SCWT is performed using the built-in MATLAB function~\cite{MATLAB2019a} and the ConceFT MATLAB implementation of Ref.~\cite{Daubechies2016}. Sample time-frequency representations are shown in Fig.~\ref{fig:TF} for the oscillation signal presented in Fig.~3(a) of the main text. From this time-frequency representation, we then extract the peak frequency and width (as a measure of error) as a function of time as shown in Fig.~4(a) of the main text.

\begin{figure}[h]
	\includegraphics[width=\columnwidth]{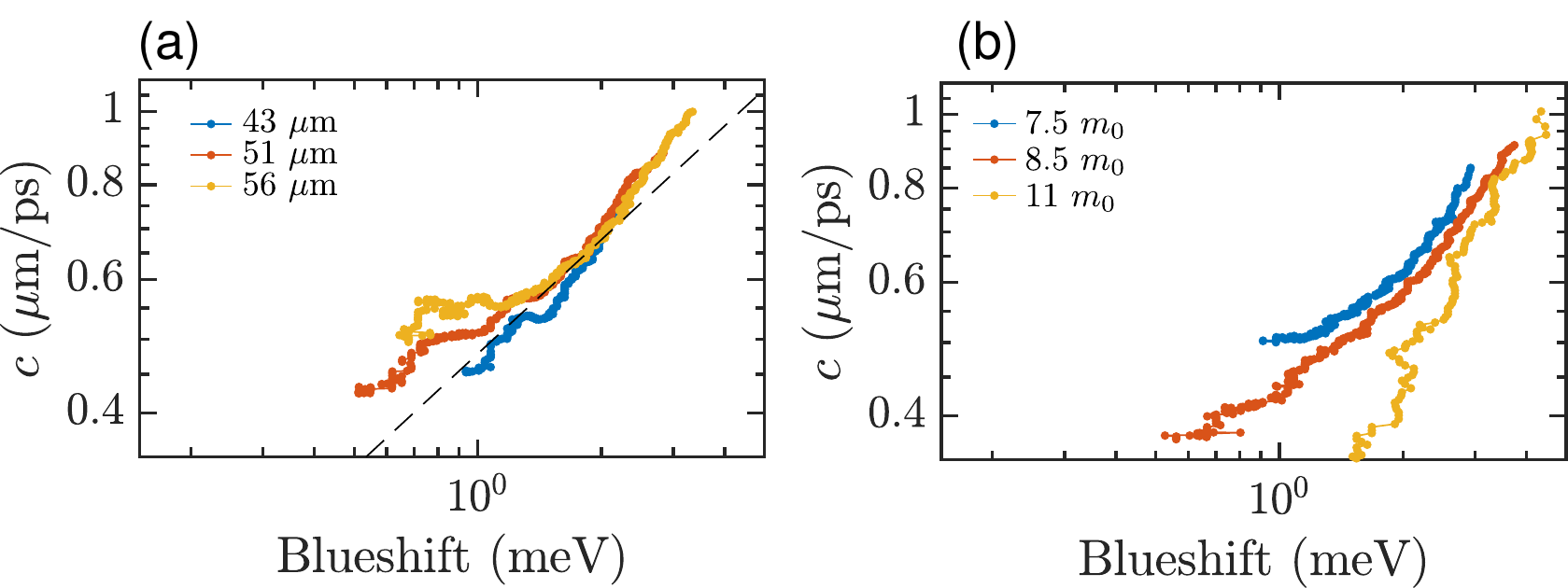}
	\caption{\label{fig:size} Extracted speed of sound for different (a) trap diameters and (b) polariton effective masses.}
\end{figure}

\begin{figure}[th]
	\includegraphics[width=\columnwidth]{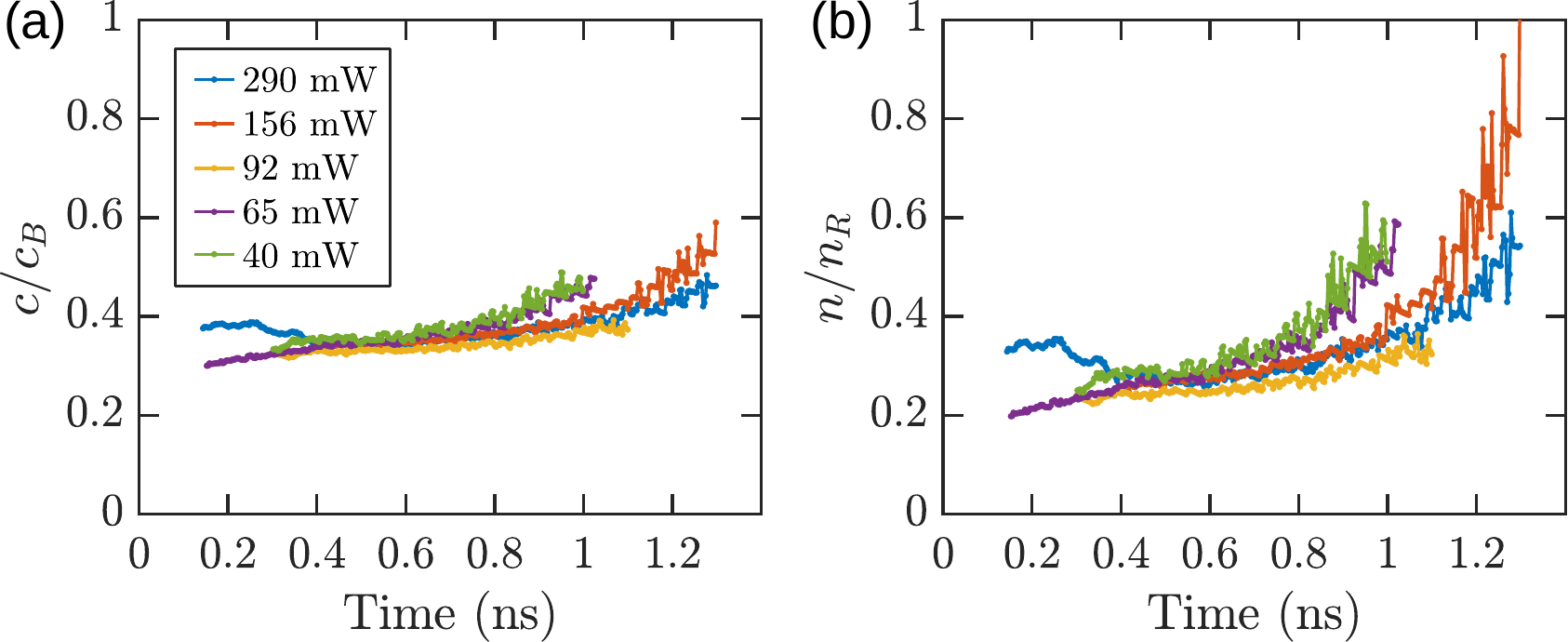}
	\caption{\label{fig:reservoirRatio} (a) Time-dependence of the measured speed of sound $c$ relative to the Bogoliubov sound $c_B$ as a function of time. (b) Estimated polariton--reservoir density ratio $n/n_R$ assuming $c=\sqrt{gn/m}$, as a function of time. All data correspond to the data presented in Fig.4(b) of the main text.}
\end{figure}

\textbf{Dependence on the ring size and effective mass:}
We also performed the same analysis for different trap sizes and effective masses (measured at different detunings) as shown in Fig.~\ref{fig:size}.
The behavior follows the results presented in Fig.~4(c) of the main text where the curves collapse to the square-root-like behavior at larger blueshifts, independently of the trap size, as shown in Fig.~\ref{fig:size}(a). However, at smaller blueshifts, the speed of sound is larger for larger rings. This can be attributed to the non-negligible density of the reservoir particles inside the ring which we expect to be smaller in larger rings. For the same measured blueshift, a smaller reservoir density means a higher polariton-polariton interaction energy, hence a higher measured speed of sound (see Fig.~\ref{fig:reservoirRatio}). The behavior is also consistent across a small range of polariton effective masses (or different detunings) tested in the experiment as shown in Fig.~\ref{fig:size}(b).
Note that the ranges of trap sizes and exciton-photon detunings accessible in the experiment are limited.
Namely, the oscillation signal is below the noise level for smaller traps ($D<40~\mu$m) but the pump power is not sufficient for high-density excitation of larger traps ($D>60~\mu$m).
The detuning range is limited to near zero detuning, i.e. near-resonance of the cavity photon and quantum well exciton. For negative detunings, a higher excitation power is needed to create a single-mode condensate in the Thomas-Fermi regime, while for positive detunings, the underlying optically induced potential strongly deviates from a box-like shape, as we have previously shown in Ref.~\cite{Estrecho2019}.
Nevertheless, the small effective mass range shown in Fig.~\ref{fig:size}(a) indicates that the measured speed of sound decreases with increasing mass, as expected.

\textbf{Estimation of the polariton--reservoir ratio:}
The ratio of the measured speed of sound $c$ to the the Bogoliubov sound $c_B=\sqrt{gn/m}$ is presented in Fig.~\ref{fig:reservoirRatio}(a). To arrive to this result, the first step is to convert the extracted frequency vs time to the speed of sound vs time. Using the one-to-one correspondence of the time to the blueshift presented in Fig. 1(b) of the main text, we can convert time to energy and extract the speed of sound vs the blueshift, as shown in Fig. 4(c) of the main text. Finally, we normalize the measured speed of sound to the Bogoliubov sound (assuming that the blueshift comes from the polariton-polariton interactions in the condensate), but revert to the time axis.

At early times, $c/c_B\approx 1/3$ and rises towards $c/c_B\approx 1/2$ at later times. Assuming that the measured speed of sound is the Bogoliubov sound, i.e. $c=c_B$, we can estimate the polariton--reservoir ratio~\cite{Stepanov2019} inside the trap as follows:
\begin{gather*}
\Delta E = gn + |X|^{-2}gn_R = mc^2 + |X|^{-2}gn_R \\
\frac{n}{n_R} = |X|^{-2}\left( \frac{\Delta E}{mc^2} - 1 \right)^{-1}
\end{gather*}
where $|X|^{2} \approx 0.5$ at near zero detuning. The time-dependence of this ratio is shown in Fig.~\ref{fig:reservoirRatio}(b) for the experimental data presented in Fig.~4 of the main text.

\bibliographystyle{apsrev4-2}